\documentclass[12pt,superscriptaddress,lineno]{revtex4-2}

\usepackage[english]{babel}
\usepackage[utf8]{inputenc}

\usepackage{graphicx}

\usepackage{amsmath}
\usepackage{amssymb}
\usepackage{amsthm}
\usepackage{nicefrac}
\usepackage{lineno}
\usepackage{xcolor}   
\usepackage{verbatim}
\usepackage{ulem}
\usepackage{svg}
\usepackage{siunitx}

\usepackage[colorlinks=true,
            citecolor=blue,
            linkcolor=black,
            anchorcolor=black,
            urlcolor=blue]{hyperref}

\begin{document}
\nolinenumbers   
\title{Fast Nuclear Spin Read-Out in Single Molecule Magnets}
\author{Hongyan Chen}
    \thanks{H.C. and S.G. contributed equally to this work.}
    \email{chenhongyan@ziac.cn}
    \altaffiliation[Present address: ]{Zhongshan Institute of Advanced Cryogenic Technology, Zhongshan 528400, China}
	\affiliation{Physikalisches Institut (PHI), Karlsruhe Institute of Technology (KIT), 76131 Karlsruhe, Germany} 
\author{Simon Gerber}
    \thanks{H.C. and S.G. contributed equally to this work.}
    \email{simon.gerber@kit.edu}
	\affiliation{Physikalisches Institut (PHI), Karlsruhe Institute of Technology (KIT), 76131 Karlsruhe, Germany}
\author{Charanpreet Singh}
    \email{charanpreet.singh@kit.edu}
	\affiliation{Physikalisches Institut (PHI), Karlsruhe Institute of Technology (KIT), 76131 Karlsruhe, Germany}
\author{Nola Warwick}
	\affiliation{Physikalisches Institut (PHI), Karlsruhe Institute of Technology (KIT), 76131 Karlsruhe, Germany}
\author{Philip Schmid}
	\affiliation{Physikalisches Institut (PHI), Karlsruhe Institute of Technology (KIT), 76131 Karlsruhe, Germany}
\author{Svetlana Klyatskaya}
	\affiliation{Institute of Nanotechnology (INT), Karlsruhe Institute of Technology (KIT), 76344 Eggenstein-Leopoldshafen, Germany}	
\author{Eufemio Moreno-Pineda}
	\affiliation{Institute for Quantum Materials and Technologies (IQMT), Karlsruhe Institute of Technology (KIT), Hermann-von-Helmholtz-Platz 1, D-76344, Eggenstein-Leopoldshafen, Germany}
    \affiliation{Universidad de Panamá, Facultad de Ciencias Naturales, Exactas y Tecnología, Depto. de Química-Física, 0824 Panamá, Panamá}
\author{Mario Ruben}
	\affiliation{Institute for Quantum Materials and Technologies (IQMT), Karlsruhe Institute of Technology (KIT), Hermann-von-Helmholtz-Platz 1, D-76344, Eggenstein-Leopoldshafen, Germany}
	\affiliation{Institute of Nanotechnology (INT), Karlsruhe Institute of Technology (KIT), 76344 Eggenstein-Leopoldshafen, Germany}
	\affiliation{Centre Européen de Sciences Quantiques (CESQ) in the Institut de Science et d'Ingénierie Supramoléculaires (ISIS), 8 allée Gaspard Monge BP 70028, 67083 Strasbourg Cedex France}
\author{Wulf Wulfhekel}
	\affiliation{Physikalisches Institut (PHI), Karlsruhe Institute of Technology (KIT), 76131 Karlsruhe, Germany}
	\affiliation{Institute for Quantum Materials and Technologies (IQMT), Karlsruhe Institute of Technology (KIT), Hermann-von-Helmholtz-Platz 1, D-76344, Eggenstein-Leopoldshafen, Germany}

\date{\today}

\begin{abstract}
Nuclear spins in $4f$ single-molecule magnets (SMMs) are promising qubits or qudits candidates for quantum information processing due to their relative isolation and reduced susceptibility to environmental disturbances, while their hyperfine coupling to the $4f$ magnetic moments enable readout and control. 
So far, the nuclear spin states of individual TbPc$_2$ SMMs have been detected in transport measurements via the spin-cascade effect.
Here, we demonstrate a fast, field-sweep-free readout of nuclear spin in individual $^{163}$DyPc$_2$ ($I=5/2$) and $^{159}$TbPc$_2$ ($I=3/2$) using millikelvin spin-polarized scanning tunneling microscopy. 
This fast readout is enabled by the hyperfine interactions, which modifies the statistics of the telegraph noise generated by reversals of the Ln$^{3+}$ moment, thereby revealing the nuclear spin state.
We observe nuclear spin relaxation times $T_1$ in excess of minutes. 
Furthermore, we drive nuclear spin transitions using a radio-frequency field and detect the resulting nuclear magnetic resonance directly in the tunneling current.
Thus our approach enables fast access to long-lived nuclear spin states in an individual molecule.
\end{abstract}

\maketitle
\section{Main}
Quantum information processing with single spins requires conflicting conditions, i.e., the quantum spin should be isolated from the bath to ensure long lifetimes and coherence, while still allowing control and readout via external stimuli \cite{atzori_second_2019,moreno-pineda_molecular_2018,wolfgang2019,goodwin2017molecular,shiddiq2016enhancing,TbPc2PhysRevLett.107.177205,thiele2014electrically,thiele2013electrical,Ln-SMM2013review,RE-SMM80Khysteresis2018,Rf-STSforTbPc25K}.
A particularly successful approach is the so-called spin cascade in metal-organic molecules, in which the qubit or qudit (a quantum system with more than two states) is encoded in the nuclear spin $I$ of a rare-earth ion \cite{thiele2013electrical}.
The coupling to the external world is mediated by the hyperfine interaction with the electronic total angular momentum $J$ of the rare-earth ion, which is in turn exchange-coupled to a single electron spin $S$ on the organic ligands of a phthalocyanine (Pc) double-decker molecule \cite{vincent2012electronic}.
This unpaired electron undergoes Kondo screening when the molecule is placed on a conductive surface \cite{komeda2011observation, KOMEDA2014127, DyPc2onCu001_2016}.
In Au break-junctions, this mechanism has been used to read, write and manipulate the nuclear spin of TbPc$_2$ \cite{vincent2012electronic,thiele2013electrical} and the Grover's algorithm has been realized \cite{PhysRevLett.79.325,leuenberger_quantum_2001,PhysRevLett.119.187702}. 
To read out the nuclear spin, the gate voltage was tuned to the Kondo resonance while sweeping the magnetic field normal to the Pc ligands. This drives a reversal of the ion's total angular momentum at the avoided level crossings of the hyperfine-split states \cite{vincent2012electronic,thiele2013electrical}.
As the magnetic field for the avoided level crossing depends on the nuclear spin state, the reversal field reveals the nuclear spin state.
The measurement therefore constitutes a projective measurement of the nuclear spin.

\begin{figure*}
  \centering
  \includegraphics[width=\textwidth]{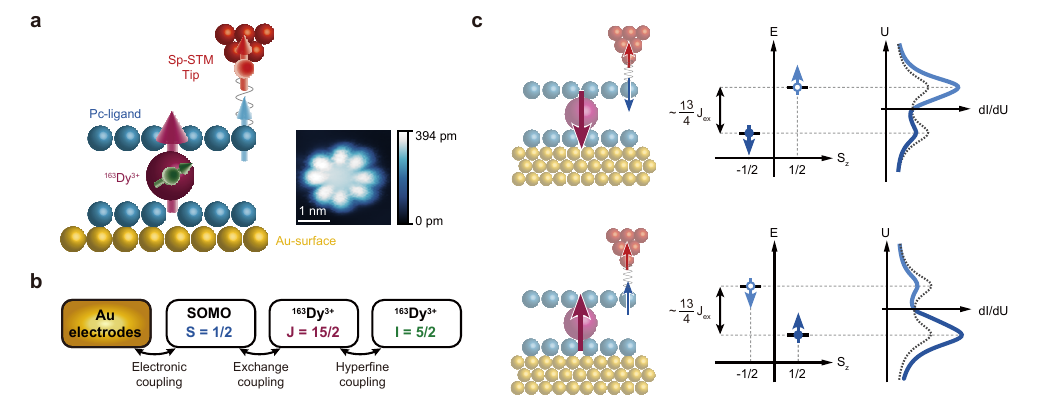}
  \caption{Spin-polarized STM setup for readout principle based on the spin cascade. (a) Schematic side view of $^{163}$DyPc$_2$ molecule on Au(111) measured using a spin-polarized tip (red arrow). The ligand spin $S$ (blue arrow) is exchange-coupled to the $^{163}$Dy$^{3+}$ total angular momentum $J$ (purple arrow), which in turn, is hyperfine-coupled to the nuclear spin $I$ (green arrow). Inset: STM topography of a $^{163}$DyPc$_2$ molecule on Au(111) acquired at \SI{1}{V}, \SI{20}{pA}. (b) Diagrammatic representation of the resulting spin cascade, illustrating the electronic, exchange, and hyperfine couplings. (c) Spin-polarized STM readout of the $^{163}$Dy$^{3+}$ moment. The top and bottom rows represents the two opposite orientation associated with $J_z =\pm 13/2$. The middle part of both rows illustrate the degeneracy lifting of ligand spin $S_z=\pm 1/2$ by $13J_{\mathrm{ex}}/4$. For fixed spin polarization of the tip, the resulting tunneling spectra for two cases is shown on the right by the blue line. Dotted line represent the spectra for a non-polarized tip.}
  \label{Fig1}
\end{figure*}

In this work, we take an alternative approach using the same spin cascade, but without the need to sweep an external magnetic field, thus allowing a faster readout of the nuclear spin states.
We demonstrate this scheme for $^{163}$DyPc$_2$ and TbPc$_2$ single-molecule magnets using spin-polarized scanning tunneling microscopy (Sp-STM) at around \SI{35}{mK}.
The basic idea of the scheme is illustrated in Fig. \ref{Fig1}. 
$^{163}$DyPc$_2$ and TbPc$_2$ molecules were deposited onto a clean Au(111) surface by sublimation \cite{moreno2017nuclear,JACS1980,InorgChem1988,koike1996relationship,ishikawa2003determination,ishikawa2003lanthanide}. 
In STM scans, they exhibit the characteristic cloverleaf structure of the upper Pc ligand (Fig. \ref{Fig1}a) \cite{komeda2011observation,frauhammer_indirect_2021}. 
Owing to the energy-level alignment with the Au(111) work function, the upper ligand hosts an unpaired electron forming a singly occupied molecular orbital (SOMO) \cite{vrubel_ab_2020,KOMEDA2014127,frauhammer_indirect_2021}. 
This unpaired electron carrying spin $S=1/2$ is Kondo-screened by the substrate electrons and is exchange-coupled to the moment of the Ln$^{3+}$ ion.
This exchange interaction acts as an effective magnetic field on the ligand spin, splitting the Kondo resonance \cite{vincent2012electronic,frauhammer_indirect_2021} into two spin-polarized peaks and thereby encoding the orientation of the Ln$^{3+}$ total angular momentum ($J$) in the electronic spin state of the SOMO ($S$). 
The Ln$^{3+}$ moment is in turn hyperfine-coupled to the nuclear spin ($I$), allowing the nuclear spin state to influence the Kondo resonance via the spin cascade. 
The spin-polarized Kondo resonance is then detected by spin-polarized STM revealing the nuclear spin state. 

The first step of the spin cascade is the readout of the Ln$^{3+}$ moment, which has been demonstrated for $^{164}$DyPc$_2$ in our previous work \cite{frauhammer_indirect_2021}.
The corresponding STM measurement principle is illustrated in Fig. \ref{Fig1}c.
Note that the ground state doublet of the Dy$^{3+}$ ion ($J=15/2$) are the states  $J_z=\pm 13/2$ \cite{ishikawa2003lanthanide}.
Depending on the orientation of the Dy$^{3+}$ moment ($J_z=\pm 13/2$) the two spin states of the unpaired electron ($S_z=\pm 1/2$) are exchange-split by $\pm \frac{13}{4}J_\mathrm{ex}$, resulting in fully spin-polarized split Kondo peaks \cite{von2015spin,frauhammer_indirect_2021,Bagchi2024}.
With spin-polarized scanning tunneling spectroscopy (STS), the differential conductance varies due the tunneling magnetoresistance effect (TMR) \cite{julliere1975tunneling}, resulting in different peak intensities for the two spin split Kondo peaks.
This difference between the peak intensities depend on relative orientation Dy$^{3+}$ moment and tip polarization, hence
allows to determine the orientation of the Dy$^{3+}$ moment.

\begin{figure*}
  \centering
  \includegraphics[width=\textwidth]{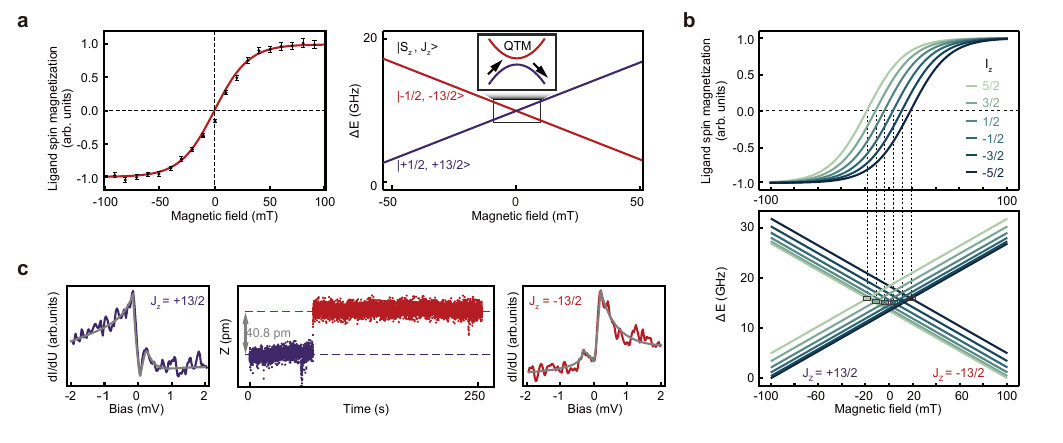}
  \caption{Effect of the spin cascade on Dy$^{3+}$ moment (a) Field-dependent magnetization data for $^{164}$DyPc$_2$ ($I=0$) measured with magnetic field along the easy axis (left), together with the corresponding Zeeman diagram (right). The red curve is calculated magnetization for a $T=\SI{200}{mK}$ which agrees well with $T_\mathrm{eff}=\SI{206}{mK}$ of the experimental data. (b) Simulated magnetization curves (top) and Zeeman diagram (bottom) for $^{163}$DyPc$_2$ with nuclear spin $I=5/2$. The different shades of green represent the different nuclear spin states $I_z$. (c) Zero-field dependent magnetic switching of an individual $^{163}$DyPc$_2$ ($I=5/2$) molecule. The middle panel shows a discrete jump in the constant-current $z(t)$ signal, with d$I$/d$U$ recorded before (left) and after (right) the jump. The spectral inversion identifies the jump as a reversal of Dy$^{3+}$ moment. }
  \label{Fig2}
\end{figure*}

\begin{figure*}
  \centering
  \includegraphics[width=\textwidth]{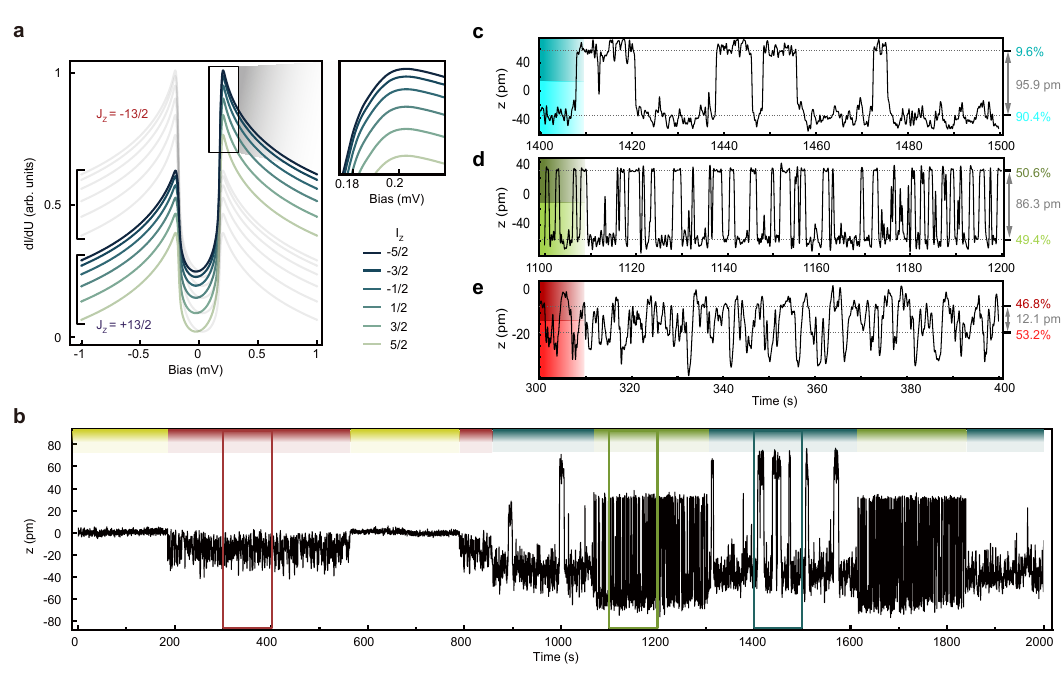}
  \caption{Nuclear-spin-dependent switching dynamics of $^{163}$DyPc$_2$ at zero field (a) Calculated spin-polarized differential conductance for the six nuclear spin states $I_z$ with a \qty{30}{\percent} spin polarization of the tip. (b) Time trace of the tip z-position on top of a $^{163}$DyPc$_2$ molecule at $B=\qty{0}{mT}$ and $U=\qty{240}{\micro\volt}$. The trace exhibits different sections with distinct switching rates and amplitudes,  grouped by different colors. (c)-(e) Zoomed in plots for the colored sections with population and jump heights marked on the right side.}
  \label{Fig3}
\end{figure*}
  
Using this principle, the magnetization curve of $^{164}$Dy$^{3+}$ without nuclear spin as a function of an external magnetic field applied normal to the surface was previously determined \cite{frauhammer_indirect_2021}.
The magnetization shows no remanence and vanishes at zero field (Fig. \ref{Fig2}a). 
This behavior originates from quantum tunneling of magnetization (QTM) at an avoided level crossing between the two lowest-energy states of the coupled Dy-ligand spin system, $|S_z,J_z>$ = $|+1/2, +13/2>$ and $|-1/2,-13/2>$. 
The avoided level crossing results from mixing terms of the Stevens operators describing the crystal field splitting \cite{stevens1952matrix, miyamachi2013stabilizing, balashov2018electron}. 
On top of that, as the total angular momentum of the coupled Dy–ligand system $J \otimes S$ is integer, the ground state is not protected by Kramers degeneracy.

\begin{figure*}
  \centering
  \includegraphics[width=\textwidth]{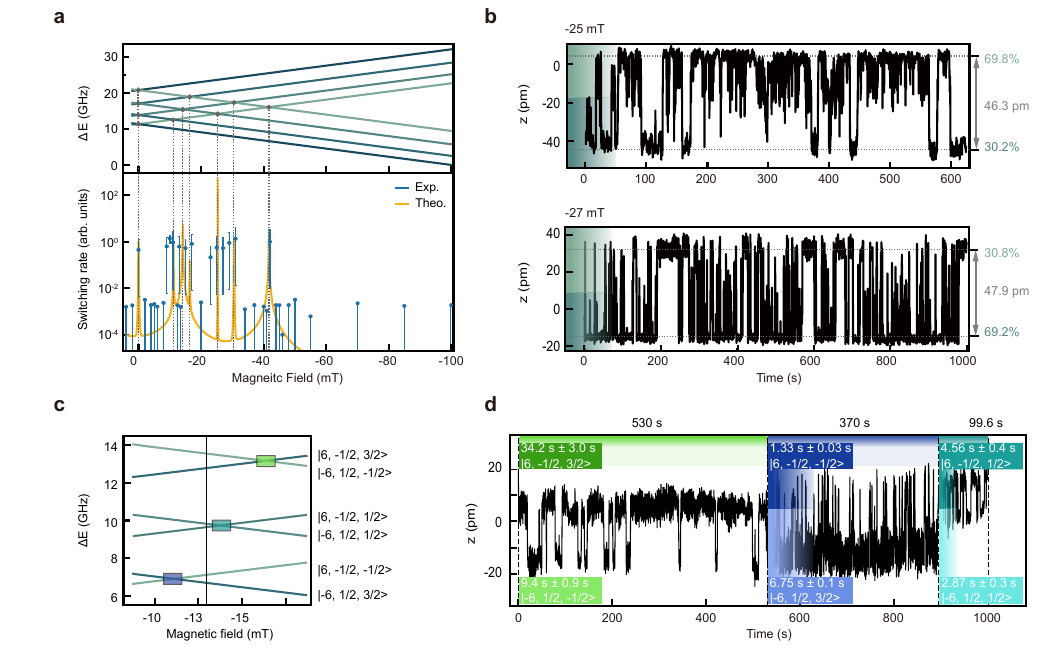}
  \caption{Field-dependent RTN of TbPc$_2$ (a) Comparison between Zeeman diagram and magnetic field dependent switching rates of Tb$^{3+}$ moment. The orange line represents the calculated switching rates due to the tunneling current whereas the blue markers show the experimentally measured rates. Enhanced switching occurs at magnetic fields close to the calculated avoided level crossings, marked by dotted vertical lines. (b) Time traces acquired at \qty{-25}{\milli\tesla} (top) and \qty{-27}{\milli\tesla} (bottom) on opposite sites of an avoided level crossing. The population asymmetry reverses around an avoided level crossing. (c) Magnified Zeeman diagram with three avoided level crossings in close proximity to the magnetic field of the measurement (black line). Colored markers denote the QTM points associated with different section shown in d. (d) RTN signal recorded at magnetic field indicated in c, showing three representative sections with distinct switching rates. }
  \label{Fig4}
\end{figure*}
The situation changes qualitatively for $^{163}$DyPc$_2$ where the $^{163}$Dy isotope carries a half integer nuclear spin $I=5/2$.
The hyperfine interaction between the Dy$^{3+}$ total angular momentum ($J$) and the nuclear spin splits the energies of the electronic states depending on the nuclear spin projection as depicted in Fig. \ref{Fig2}b.
As a result, the magnetization curve of the Dy$^{3+}$ ion is shifted along magnetic field axis for different nuclear spin states, leading to finite expectation values of the Dy$^{3+}$ magnetization even at zero external magnetic field.
This is in contrast to $^{164}$DyPc$_2$.
The combined ligand-Dy-nuclear spin system $S\otimes J\otimes I$ is Kramers protected, which suppresses any QTM at zero magnetic field. 
Consistent with this picture, a finite Dy$^{3+}$ magnetization at zero magnetic field was experimentally observed, as shown in Fig. \ref{Fig2}c. 
The left panel shows the spin-polarized STS signal where the observed asymmetry indicates a stable Dy$^{3+}$ moment. 
Parking the tip above the molecule and recording a time trace of the $z$-position, a jump was detected and spectroscopy after the jump revealed an inversion of asymmetry, that is a flip of the Dy$^{3+}$ moment.
Thus the Dy$^{3+}$ moment for $^{163}$DyPc$_2$ remains stable for minutes in marked contrast to the zero-field instability to $^{164}$DyPc$_2$.
The Dy$^{3+}$ moment remains stable for minutes and exhibit a clear spin polarization of the split Kondo peaks.

To better understand the microscopic origin of this behavior, the effective spin Hamiltonian of the $^{163}$DyPc$_2$ molecule is considered.
\begin{equation}
\begin{split}
    \hat{H}= & B_2^0 \hat{O}_2^0 + B_2^2 \hat{O}_2^2 + B_4^0 \hat{O}_4^0 + B_4^4 \hat{O}_4^4 + B_6^0 \hat{O}_6^0 + B_6^4 \hat{O}_6^4 \\
    & + A_{IJ}  \hat{\mathbf{I}} \cdot \hat{\mathbf{J}}\\
    &+P \{ \hat{I}_z^2-\frac{1}{3}I(I+1) \} \\
    &+ g_{\mathrm{J}} \mu_{\mathrm{B}} \hat{\mathbf{J}}\cdot \mathbf{B} + g_{\mathrm{S}}\mu_{\mathrm{B}} \hat{\mathbf{S}}\cdot \mathbf{B}+ g_{\mathrm{I}}\mu_{\mathrm{B}}\hat{\mathbf{I}} \cdot\mathbf{B}\\
    &-J_{\mathrm{ex}} \hat{\mathbf{J}}\cdot\hat{\mathbf{S}}    
\end{split}
\label{equ1}
\end{equation}
\noindent Here, $\hat{\mathbf{S}}$ denotes the ligand spin operator with spin quantum number $S=1/2$, $\hat{\mathbf{J}}$ the Dy$^{3+}$ total angular momentum operator with $J=15/2$, and $\hat{\mathbf{I}}$ the nuclear spin operator of $^{163}$Dy with $I=5/2$. 
$J_{\mathrm{ex}}$ describes the exchange interaction between $\hat{\mathbf{S}}$ and $\hat{\mathbf{J}}$. $\hat{O}_n^m$ and $B_n^m$ are the Stevens operators and the corresponding crystal-field parameters of DyPc$_2$ taken from Ref. \cite{ishikawa2005quantum, ishikawa2003determination}. 
The parameters reflect a DyPc$_2$ molecule of C$_{\text{2v}}$ symmetry.
$A$ and $P$ denote the hyperfine and quadrupole parameters of $^{163}$DyPc$_2$ \cite{moreno2017nuclear} (see Supplemental Material).

Using this Hamiltonian, the Zeeman diagram of the coupled ligand-Dy-nuclear spin system around zero magnetic field is calculated, shown in Fig. \ref{Fig2}b.
The two ground-state branches, $J_z=\pm 13/2$, are each split into six different sublevels by the hyperfine coupling to the nuclear spin $I=5/2$. 
Around zero magnetic field, the eigenstates remain essentially unmixed and are nearly pure in the $z$ components of the individual spins, since the exchange and quadrupole terms do not mix the corresponding $z$-eigenstates.
Only a minute mixing arises from the non-diagonal crystal-field terms $\hat{O}_2^2$ and $\hat{O}_4^4$, which also generate avoided level crossings at finite fields.
While in the process of Kondo screening and the measurement with the STM, the ligand spin flips at a high rate.
At the same time, the Ising nature of the giant Dy spin with its essentially pure $J_z$ states prevents the reversal of $J$. 
This also screens the nuclear spin from spin flips by the conduction electrons. 
This energy level structure explains the stability of the Dy$^{3+}$ magnetization observed at zero field in Fig. \ref{Fig2}c. 
When, however, the magnetic field is tuned closer to any avoided level crossing of the hyperfine levels, the states are not anymore pure in the $z$-components of the spins, and therefore $J$ as well as $I$ are not protected from spin flips by the fluctuating ligand spin $S$. 
\begin{figure*}
  \centering
  \includegraphics[width=\textwidth]{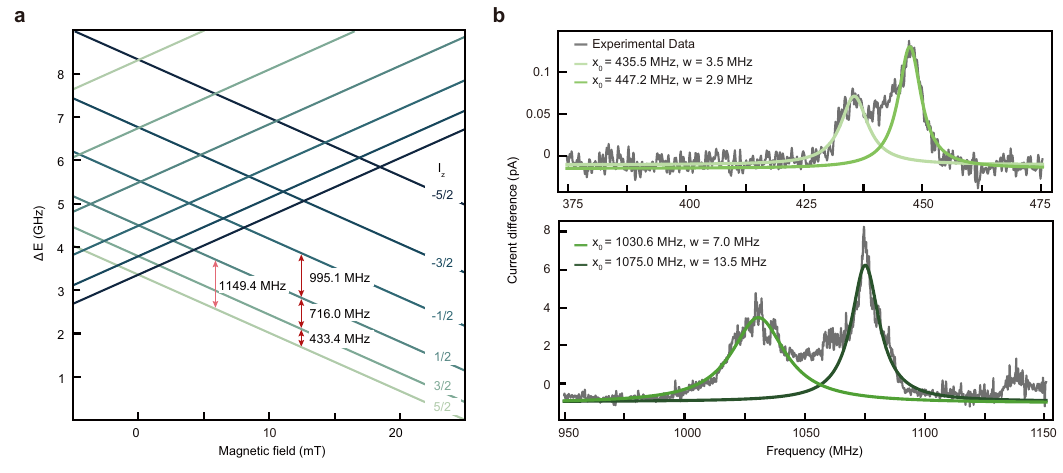}
  \caption{Radio-frequency excitation of nuclear spin states in $^{163}$DyPc$_2$ (a) Zeeman diagram of $^{163}$DyPc$_2$ with double headed errors marking the calculated transition frequency for dipole (red) and quadrupole (pink) direct transitions. (b) Change in tunneling current as a function of radio-frequency. The top panel shows the frequency range corresponding to the lowest dipole transition where as the lower panel shows lowest quadrupole transition. }
  \label{Fig5}
\end{figure*}

The splitting of the Kondo peak originates from the exchange interaction between the Kondo-screened ligand spin and the Dy$^{3+}$ moment. 
As the energies of the states involved in Kondo screening are further modified by the hyperfine interaction with the nuclear spin, the height of the Kondo peaks can therefore depend on the nuclear spin state.
This potentially provides a direct spectroscopic pathway to probe the nuclear spin state.

To quantify this effect, the differential conductance $dI/dU$ is calculated using the third-order perturbation model developed by Ternes for spin-flip scattering in spin-polarized STM \cite{ternes2015spin}. 
In this approach the transition probabilities for tunneling electrons to induce changes of the ligand spin $S$, the Dy$^{3+}$ total angular momentum $J$ and the nuclear spin $I$ were computed.
The calculated transition probabilities at zero applied magnetic field show that processes involving changes of the nuclear spin are seven orders of magnitude less probable that spin flips of the ligand spin for first order processes. 
In the calculations of the Kondo peak, they are therefore irrelevant on the level of the present third-order treatment.
This suppression arises because the eigenstates are nearly pure in the individual spin projections, i.e., the corresponding spin operators almost commute with the Hamiltonian. 
Consequently, the tunneling interaction induces almost exclusively spin-flip transitions of the ligand spin.
The Dy$^{3+}$ moment and the nuclear spin therefore remain effectively stable and do not scatter with conduction electrons, in agreement with the single-molecule transport measurements \cite{vincent2012electronic}.
The simulations of the Kondo spectra were thus based on spin transitions of the ligand spin alone. 

Figure \ref{Fig3}a shows the calculated spin-polarized differential conductance $dI/dU$ for different nuclear spin states $I_z$.  
As can be seen from the figure, the characteristic exchange split Kondo resonance is retained, however the peak positions and intensities are modified depending on the nuclear spin state.
The energy shift originates from the small hyperfine and dipolar fields acting on the Dy$^{3+}$ moment, while the change in peak intensities arises because the energy of the localized spin in the Kondo model, i.e. the Anderson impurity states, depends on the nuclear spin projection. 
This dependence leads to a change in the local density of states (LDOS) of the Kondo resonance. 
For details of the calculations, we refer to the Supplementary Material.

To probe these nuclear-spin-dependent spectral features experimentally, spin-polarized STM measurements are performed at a bias voltage near one of the maxima of the split Kondo peak, where the expected spectral differences predicted in Fig. \ref{Fig3}a are most pronounced. 
Under constant current feedback at such a bias voltage, only the states between the Fermi level and the bias voltage contribute to tunneling.
A change of the LDOS or its spin-polarization in this energy window induces a change in the $z$-position of the STM tip in the constant current feedback mode.
The change due to spin-polarization can be clearly seen from the time trace shown in Fig. \ref{Fig2}c, where the tip position moves by several tens of pm upon flipping of the Dy$^{3+}$ moment. 
The height difference depends on the spin polarization of the STM tip.
Changes of the nuclear spin state are expected to induce smaller changes, as they do not affect the spin-polarization of the split Kondo peak.

Figure \ref{Fig3}b shows a typical time trace of the STM tip height measured on the Pc ligand at a bias voltage of $\SI{240}{\mu V}$ near the maximum of the split Kondo peak, and at vanishing external magnetic field. 
Remarkably, the time trace not only show the expected switching between two levels corresponding to the flipping of Dy$^{3+}$ moment, but rather separates into different sections, with markedly different  switching rates, level population and height. 
Representative examples are highlighted by the colored sections in Fig. \ref{Fig3}b and shown in detail in Fig. \ref{Fig3}c-e.

Each section here represents the flipping of the Dy$^{3+}$ moment, albeit at a different rate. 
The yellow section corresponds to a very stable configuration in which no spin flips occur, evident by the fact that the fluctuation level is comparable to that of background noise of STM. 
Such a scenario corresponds to the case where the avoided level crossing lies far from the applied magnetic field and the Dy moment reversal is strongly suppressed. 
For the blue section (Fig. \ref{Fig3}c), clear jumps are observed between two levels separated by approx \qty{96}{\pico\meter}, in line with the jump height observed in Fig. \ref{Fig2}c for flipping of Dy moment. 
The switching rates are expected to increase when the nuclear spin state changes bringing an avoided level crossing closer to the applied magnetic field, also resulting in a more symmetric population of the two states.
The switching becomes more frequent in the green section (Fig. \ref{Fig3}d), where the populations of the two states become nearly balanced. 
In the red section (Fig. \ref{Fig3}e), the switching rate is highest and approaches the bandwidth limit of the feedback loop, indicating that the system is closest to the avoided level crossing.
The complete behavior including switching rate and population can be explained by nuclear spin diffusion bringing the system closer or further from the avoided level crossings. 
Transitions between these sections occur on much longer time scales (several minutes) than the Dy moment switching (seconds), consistent with the long lifetime of the nuclear spin states. 
Quantitative lifetimes, decay statistics and populations are provided in the Supplementary Material. 

Depending on the nuclear spin state, the system is thus at different distance (in magnetic field) from the avoided level crossings, giving rise to characteristic switching rates of the Dy$^{3+}$ moment.
These switching rates therefore allow to determine the nuclear spin state. 
In addition, the nuclear spin state changes the Kondo peak intensity. This in principle allows a direct measurement of the nuclear spin state from the $z$-position or the current.  
This effect can be seen directly in the time trace evidenced from different $z$-positions in the four sections.
However, for DyPc$_2$, several avoided level crossings lie so close in magnetic field that we refrain to make a full determination of the nuclear quantum state.

A similar measurement for TbPc$_2$ is, however, less complex due to the smaller number of nuclear states ($I = 3/2$). 
The Tb$^{3+}$ ion has $J=6$ and the ground state doublet $J_z=\pm 6$ is separated from the next higher state by approximately \qty{600}{\kelvin} \cite{ishikawa2005quantum}.
Reversal of the Tb$^{3+}$ moment becomes possible near the avoided level crossing \cite{vincent2012electronic}, through both QTM and transitions induced by tunneling current.

At the avoided level crossings, the quantum states are mixed, the $z$-components of the individual operators $S_z, J_z$ and $I_z$ are not pure.
Consequently, the current can induce transitions via all three components of the spin operator \cite{Troiani2017,balashov2018electron}. 
This allows transitions that reverse only the Tb$^{3+}$ moment or change both the Tb$^{3+}$ moment and the nuclear spin.
As illustrated in Fig. \ref{Fig4}a, the Zeeman diagram of TbPc$_2$ has been calculated using a similar effective Hamiltonian to that of DyPc$_2$ (see Supplementary Material). 
Using the Hamiltonian, we calculate the tunneling-induced transition matrix and the switching rates are displayed in orange in the lower panel. 
The calculations yield high switching rates near avoided level crossings, indicated by the gray dotted lines in the Zeeman diagram. 
Measurements of the random telegraph noise were carried out as function of magnetic field. 
The switching rates, shown as blue points including their statistical $1\sigma$ error, match well with the simulated switching rates, thereby reproducing the position of the expected avoided level crossings. 
The experimental switching rates, however, show broader peaks indicating that additional switching processes may contribute.
At fields above \qty{50}{\milli\tesla}, we did not observe any switching events in the observation time window of \qty{20}{\minute}, setting the statistical lower cutoff.  

A further signature of the avoided level crossing is provided by the level population.
Figure \ref{Fig4}b shows time traces acquired at \qty{-25}{\milli\tesla} and \qty{-27}{\milli\tesla}, symmetrically around an expected avoided level crossing. 
Clearly, one notices that the asymmetry in the state population of the traces changes sign. 
This is expected from the level diagram that favors opposite Tb$^{3+}$ moments on either side of the avoided level crossing. 
Ultimately, we demonstrate the effectiveness of our method in identifying the nuclear spin state of the quantum system by recording time traces at a fixed magnetic field.
Figure \ref{Fig4}c-d show a magnified portion of the Zeeman diagram for TbPc$_2$, and a time trace recorded at magnetic field as indicated by the solid vertical line in panel c. 
Following the same procedure as for DyPc$_2$, we identify three different sections with distinct switching rates. 
As the switching rates reaches a maximal at a crossing point, the fastest switching section must thus lie closest to a QTM point. 
It can be identified as switching between the $|J_z,S_z,I_z>$ = $|+6, -1/2, 1/2>$ and $|-6, +1/2, 1/2>$ states, thus identifying the nuclear spin to be in the $|1/2>$ projection. 
Similarly,  the other two sections can be mapped on to the next two closest QTM points, which involves simultaneous change of the Tb$^3+$ angular momentum $J$ and nuclear spin $I$.
Note that also the $z$-position of the two Tb states vary with the nuclear spin, as the LDOS at the Kondo peak also weakly depends on the nuclear state.

We note that although the tunneling electrons induce fluctuations of the ligand spin, they almost never flip the total angular momentum or nuclear spin of 4$f$ ion when the magnetic field is far from a QTM point. 
The measurement, thus projects the system onto a nuclear spin eigenstate without transitions and realizes a nearly ideal quantum non-demolition readout. 
In this respect, the ligand spin acts as an observer quantum spin allowing for a weak measurement.

Finally, having established the nuclear spin readout through random telegraph noise, we investigate radio-frequency (RF) manipulation of the nuclear spin states. 
We follow the approach of Thiele et al. \cite{thiele2013electrical} and apply a microwave field to the tunneling junction using a separate microwave antenna \cite{peters_resonant_2020}.
In TbPc$_2$ this approach is well established to drive nuclear spin transitions via the nuclear Stark effect. 
For $^{163}$DyPc$_2$ this, a field of \SI{50}{mT} is applied to polarize the Dy$^{3+}$ moment and suppress the telegraph noise by lifting the degeneracy of the Dy states through the Zeeman interaction (Fig. \ref{Fig5}a).
The differential conductance of the Kondo peak was then recorded with a non-magnetic tip.
This makes the measurement insensitive to the orientation of Dy moment, while retaining  sensitivity to the line shape of the Kondo peak which depends on the nuclear spin state, as discussed before.
Next, the radio-frequency is swept and the tunneling current is monitored. 
When the RF frequency is resonant with a nuclear spin transition, the Kondo peak intensity changes owing to its dependence on the nuclear spin state. 
For the ground state ($I_z=5/2$), dipole transitions at \SI{433.4}{MHz} and quadrupole transitions at \SI{1149.4}{MHz} are expected.

The measured RF response in Fig. \ref{Fig5}b shows two resonances at \SI{435.5}{MHz} and \SI{447.2}{MHz}, in good agreement with the predicted dipole transition at \SI{433.4}{MHz}.
The linewidth of the resonance is approximately \SI{3}{MHz}, comparable to values reported for TbPc$_2$ molecules in break-junction experiments \cite{godfrin2017electrical}.
The splitting into two resonances is attributed to a direct interaction between the ligand spin and the Dy$^{3+}$ nuclear spin, which produces an additional hyperfine splitting. 
At higher frequencies, we also observe two peaks near the expected quadrupole transition around \SI{1151}{MHz}. 
Note that we calibrated the microwave transmission to the STM junction and account for the frequency-dependent transmission in the analysis (see Supplemental Material).


\section{Discussion}
In conclusion, a fast approach to read the nuclear spin of $^{163}$Dy and $^{159}$Tb in LnPc$_2$ single-molecule magnets is demonstrated using spin-polarized scanning tunneling microscopy.
The hyperfine interaction enables to determine the nuclear spin state of the Ln$^{3+}$ ion through telegraph-noise measurements in a spin-polarized tunneling experiment.
This mapping of nuclear spin to the telegraph noise is mediated by the spin cascade which allows us to extract the nuclear-spin-dependent relaxation times $T_1$.
Moreover, the dependence of the intensity of the Kondo peak on $I_z$ provides a simple spectroscopic tool to infer the nuclear spin state without the need for magnetic-field sweeps. 
This direct sensitivity to the nuclear spin state also allows to observe RF induced excitations of the nuclear spin.

More generally, the concept of using a Kondo-screened spin as a probe for an exchange-coupled quantum degree of freedom may be extended to a broad class of systems.
We speculate that similar schemes can also be extended to superconducting substrates, in which the split and fully spin-polarized Kondo resonance turns into a Yu-Shiba-Rusinov (YSR) state \cite{yu1965,shiba1968,rusinov1969}. 
The YSR states also offer the advantage that they are typically sharper in energy and thus the changes in the YSR states caused by the nuclear spin state lead to even larger signals. 

\section{Methods}
Extended information on used methods is provided in Supplementary Information.
\section{Acknowledgments}
We acknowledge financial support from the Deutsche Forschungsgemeinschaft
(DFG, German Research Foundation) through the Collaborative Research Centre “4f for Future” (CRC 1573, Project 471424360) Project B2 and though Project Wu 349/17-1, as well as fruitful discussions with Wolfgang Wernsdorfer, Luca Kosche, Jörg Schmalian and Anja Metelmann.
\par
H.C. and S.G. contributed equally to this work. 

\section{Author information}
\subsection{Authors and Affilations}
\textbf{Physikalisches Institut (PHI), Karlsruhe Institute of Technology (KIT), 76131 Karlsruhe, Germany}\\
Hongyan Chen, Simon Gerber, Charanpreet Singh, Nola Warwick, Philip Schmid, Wulf Wulfhekel\\
\textbf{Institute of Nanotechnology (INT), Karlsruhe Institute of Technology (KIT), 76344 Eggenstein-Leopoldshafen, Germany}\\
Svetlana Klyatskaya, Mario Ruben\\
\textbf{Institute for Quantum Materials and Technologies (IQMT), Karlsruhe Institute of Technology (KIT), 76344 Eggenstein-Leopoldshafen, Germany}\\
Eufemio Moreno-Pineda, Mario Ruben, Wulf Wulfhekel\\
\textbf{Universidad de Panamá, Facultad de Ciencias Naturales, Exactas y Tecnología, Depto. de Química-Física, 0824 Panamá, Panamá}\\
Eufemio Moreno-Pineda\\
\textbf{Centre Européen de Sciences Quantiques (CESQ) in the Institut de Science et d'Ingénierie Supramoléculaires (ISIS), 8 allée Gaspard Monge BP 70028, 67083 Strasbourg Cedex France}\\
Mario Ruben\\
\subsection{Contributions}
H.C., S.G., C.S., N.W. and P.S. carried out the measurements for DyPc$_2$-molecules. S.G. and C.S. performed the comparison measurements on TbPc$_2$ molecules. H.C., S.G., C.S., N.W. and W.W did the simulation of the molecules. E.M, S.K. and M.R. synthesised the DyPc$_2$ and TbPc$_2$ molecules. H.C., S.G., C.S. and W.W. wrote the paper with input from all authors. W.W. and M.R. supervised the project.

\subsection{Corresponding authors}
Correspondence to Simon Gerber, Charanpreet Singh or Hongyan Chen.
\section{Experimental setup}
\begin{figure*}
  \centering
    \includegraphics[width=\textwidth]{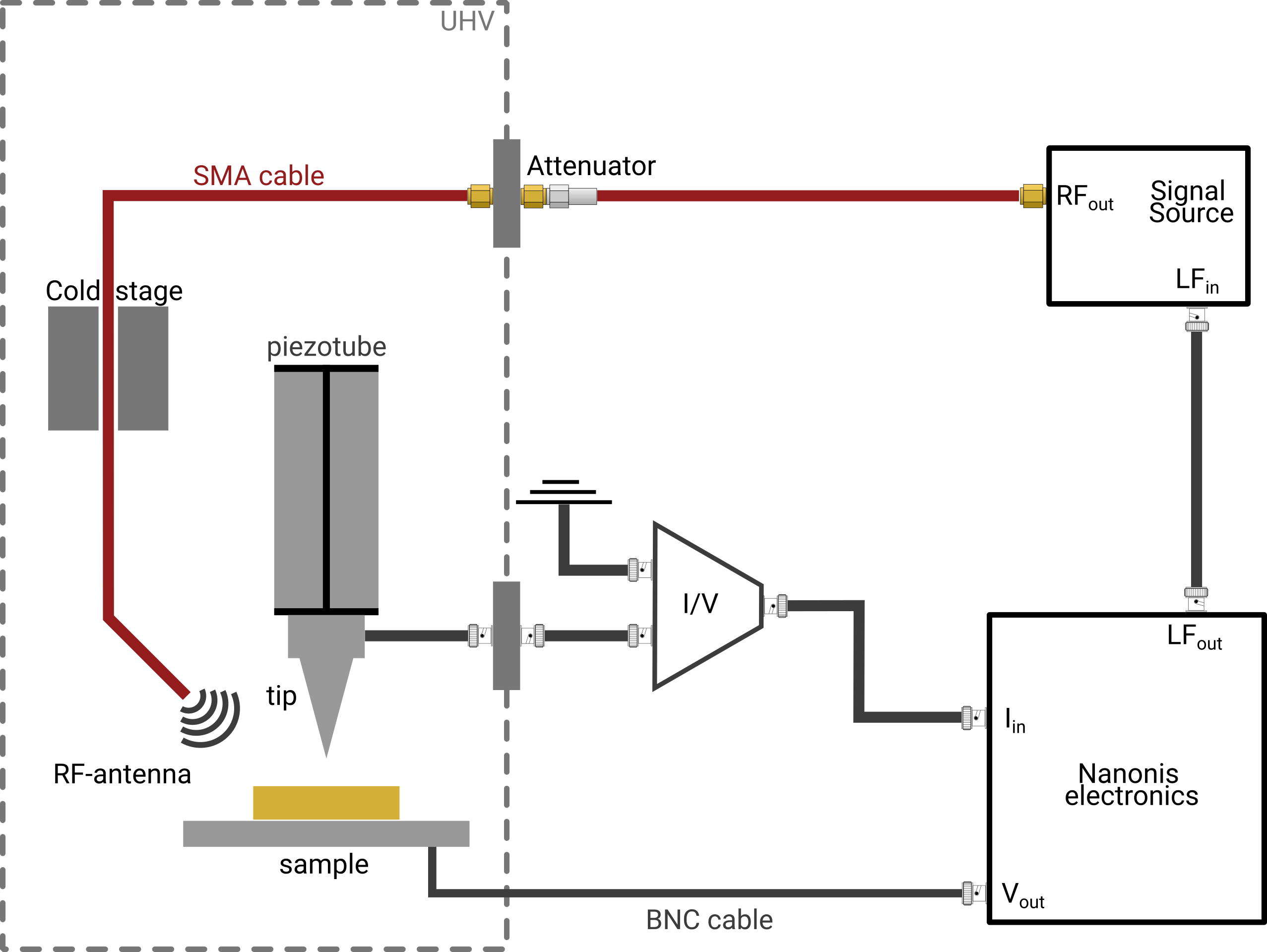}
  \caption{The configuration comprises an STM tip positioned above the sample and controlled by Nanonis electronics. For RF measurements, an RF antenna is coupled to the cryostat, thereby enabling cooling to dilution temperatures. The antenna is connected to the UHV chamber through a feedthrough and a \SI{6}{dB} attenuator. The RF signal is generated by an external source and demodulated within the Nanonis system.}
  \label{RFsetup}
\end{figure*}

The measurements were performed in a home-built ultra-high vacuum (UHV) scanning tunneling microscope (STM) that can operate at a base temperature of approximately \qty{35}{mK} and an external magnetic field of up to \qty{7}{T} can be applied normal to the surface plane \cite{balashov2018compact}.
Spin-polarized STM tips were prepared by coating a tungsten (W) tip with antiferromagnetic chromium (Cr)\cite{PhysRevLett.88.057201, PhysRevLett.92.057202}.
The out-of-plane spin polarization of the tip was adjusted by varying the thickness of the Cr coating. 
The Au(111) crystal was cleaned using repeated cycles of Ar$^+$ sputtering, followed by annealing at \qty{500}{\degreeCelsius}.
Subsequently, $^{163}$DyPc$_2$ molecules were deposited from a Knudsen cell operated at \qty{400}{\degreeCelsius} onto the clean Au(111) substrate, which was held at a temperature around \qtyrange{100}{120}{K} under UHV conditions. 
For $^{159}$TbPc$_2$ the Knudsen cell was operated at a temperature of \qty{385}{\degreeCelsius}
The substrate was then transferred in-situ to the STM.
In order to enable the control of nuclear transitions with atomic-scale spatial resolution, an RF line was integrated into the STM setup to allow for the exact measurement of the molecular response.
A schematic of the RF circuit is shown in Fig.\ref{RFsetup}.

The primary challenge in the measurement of RF induced nuclear spin transition is the transmission of the RF signal from room temperature into the UHV environment and down to dilution temperature, while avoiding additional noise, thermal heating, and mechanical vibrations.
The signal path is divided into three sections: first, a standard SMA cable with an attenuator located outside the vacuum chamber; second, an electrical feed-through; and third, a flexible cable located inside the chamber that terminates at an antenna positioned in close proximity to the tunnel junction. 
Excitation of the junction is thus achieved by the free RF field \cite{peters_resonant_2020}. 
It is evident that each transition introduces signal attenuation and reflections, phenomena that becomes more pronounced at higher frequencies.

The RF excitation is provided by an SMB 100A signal generator, while detection of small AC signals is performed using a lock-in amplifier.
The lock-in amplifier is responsible for generating the low-frequency (LF) modulation signal, that modulates the RF amplitude. 
Detection of the systems response (in the tunneling current) to the RF signal is done via the lock-in amplifier. 
The amplitude modulated RF excitation from the signal generator (\qty{100}{MHz} to \qty{2}{GHz}) is then applied to the junction via a frequency dependent transmission calibration function.

\section{Calibration of RF circuit}
The transmission calibration process involves the initial selection of an RF amplitude in the region where the measured response is linear with respect to the RF amplitude. 
Subsequently, frequency-dependent correction factors are determined in order to adjust the applied RF power and compensate for frequency dependent attenuation in the transmission line \cite{herve_rf_setup}.
The use of a \qty{6}{dB} attenuator is essential in reducing the internal reflections of signal line.
Meanwhile, the RF power applied through the transmission line must be appropriately adjusted to ensure a constant signal is received by the tunneling junction over the entire frequency range.
In order to determine the frequency-dependent response of the RF circuit, it is necessary to exploit strongly nonlinear current–voltage characteristics \cite{herve_rf_setup}.
The following reference systems have been identified as being suitable: a Pb(111) sample with sharp superconducting coherence peaks (compare \cite{peters_resonant_2020}) 
or a sharp Kondo resonance FePc molecules on Au(111) \cite{minamitani_symmetry-driven_2012}.
By monitoring the change of these nonlinear features as a function of RF frequency and amplitude, the effective RF amplitude at the junction can be calibrated.
The present procedure is designed to compensate for the intrinsic frequency dependence of the transmission line, thereby enabling quantitative comparison of frequency-dependent effects in the molecules being studied.

\begin{figure*}
    \centering
    \includegraphics[width=0.8\textwidth]{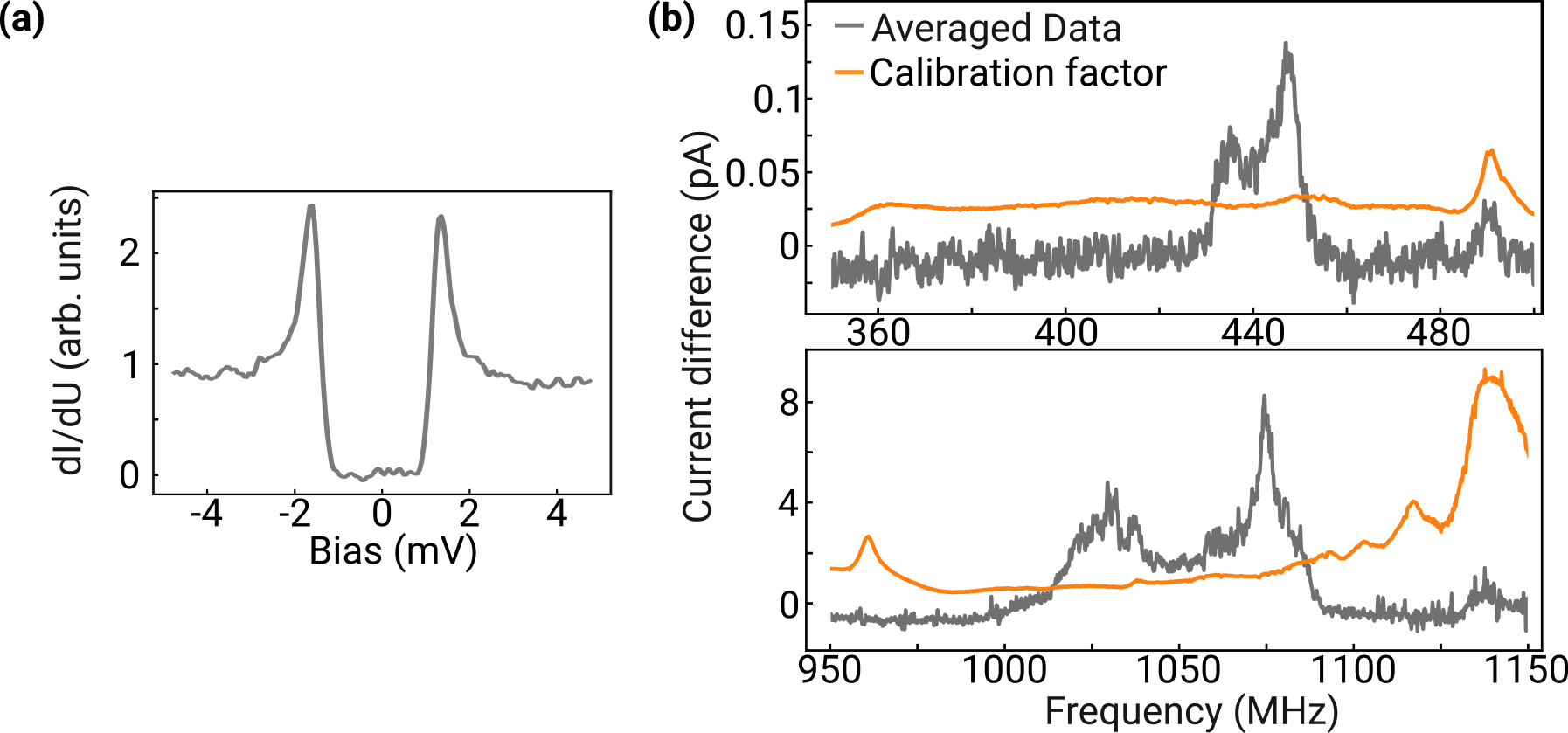}
    \caption{Transmission line calibration on a Pb(111) crystal (a) Spectroscopic measurement of the superconducting gap of a Pb(111) crystal. The coherence peaks of a superconducting gap of a Pb(111) crystal was used to calibrate our transmission line. The spectra was measured using a lock-in modulation of $U_{\mathrm{mod}}=\qty{100}{\mu V}$, feedback conditions of $U=\qty{5}{mV}, I=\qty{100}{pA}$ at a temperature of $T=\qty{800}{mK}$.(b)  A comparison of the resulting calibration factor with the independently measured transmission signal for both transitions is shown. The data in the upper panel were acquired with feedback parameters $U = \qty{150}{\mu V}, I = \qty{20}{pA}$, and an RF amplitude of $V_{RF} = \qty{350}{mV}$. The lower panel was recorded with the following feedback parameters: $U = \qty{75}{\mu V}, I = \qty{20}{pA}$, and $V_{RF} = \qty{350}{mV}$.}
    \label{trans_vs_calib}
\end{figure*}


As seen in Fig. \ref{trans_vs_calib}b, the upper panel is related to the transmission near \qty{433}{MHz}. 
In this frequency range, the signal displays a double peak around \qty{440}{MHz} and a single peak near \qty{490}{MHz}. 
Comparison with the calibration factor obtained from the Pb reference reveals that the calibration curve remains nearly constant around \qty{440}{MHz}, while it shows a peak of a similar shape at \qty{490}{MHz}.
We therefore attribute the double peak near \qty{440}{MHz} to a transitional peak. 
This is because it does not coincide with any structure in the calibration file. 
In contrast, the peak at \qty{490}{MHz} matches a feature in the calibration factor and is thus assigned to imperfect calibration rather than to an intrinsic transition.
A similar behavior is seen for the second transition near \qty{1050}{MHz} in the lower panel of the figure. 
At this frequency, the calibration factor remains constant at the position of the measured transition peaks, thereby validating their physical origin. 
However, the calibration curve displays a maximum at \qty{1130}{MHz}, which is also observed in the transmission measurement. 
Therefore, this feature is attributed to transmission-line interference rather than an intrinsic transition.

\begin{figure*}
    \centering
    \includegraphics[width=\textwidth]{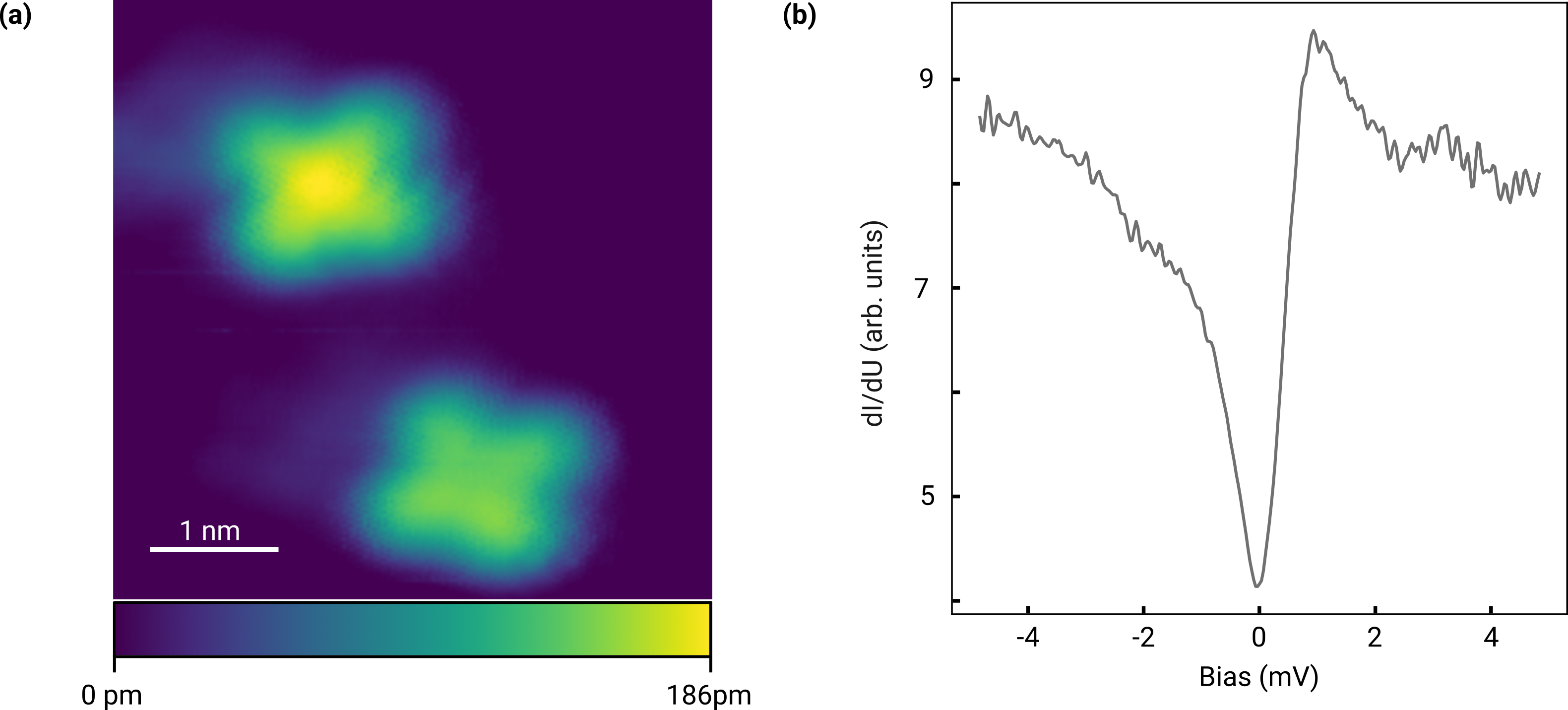}
    \caption{Scanning tunneling microscopy and scanning tunneling spectroscopy (STS) were used to investigate FePc molecules on an Au(111) surface. (a) An STM image (\qty{34}{pA}, $\qty{1}{V}$) showing two cross-shaped FePc molecules adsorbed on the Au(111) surface. (b) STS spectrum (\qty{5}{mV}, \qty{100}{\pico\ampere}, Lockin: $\qty{500}{\mu V}, \qty{3230}{Hz}$)acquired on the molecular center exhibiting a sharp dip feature near zero bias that is consistent with a Kondo resonance for on-top site FePc molecules \cite{minamitani_symmetry-driven_2012}.}
    \label{FePc molecules}
\end{figure*}

The calibration using the sharp Kondo resonance observed for FePc molecules adsorbed on Au(111) has the advantage, that no change of sample was necessary between calibration and measurement.  
For this purpose, FePc molecules were co-deposited on the Au(111) crystal together with $^{163}$DyPc$_2$, and as shown in Fig. \ref{FePc molecules}a, two cross-shaped molecules are visible, with the higher one exhibiting an apparent height of \qty{186}{pm}. 
These characteristics are attributed to FePc molecules on the surface. 
Experiments show that FePc adsorbs on Au(111) in different configurations, commonly referred to as on-top and bridge sites \cite{minamitani_symmetry-driven_2012}. 
These configurations exhibit distinct spectroscopic signatures.
The configuration associated with a sharp dip in the differential conductance spectrum, shown in Fig. \ref{FePc molecules}b, is consistent with the Kondo feature that was reported for the on-top adsorption site.
A second option for calibration was given by sharp Kondo peak from on-top site positioned FePc molecules. 
\section{Lifetimes}
\begin{figure*}
    \centering
    \includegraphics[width=\textwidth]{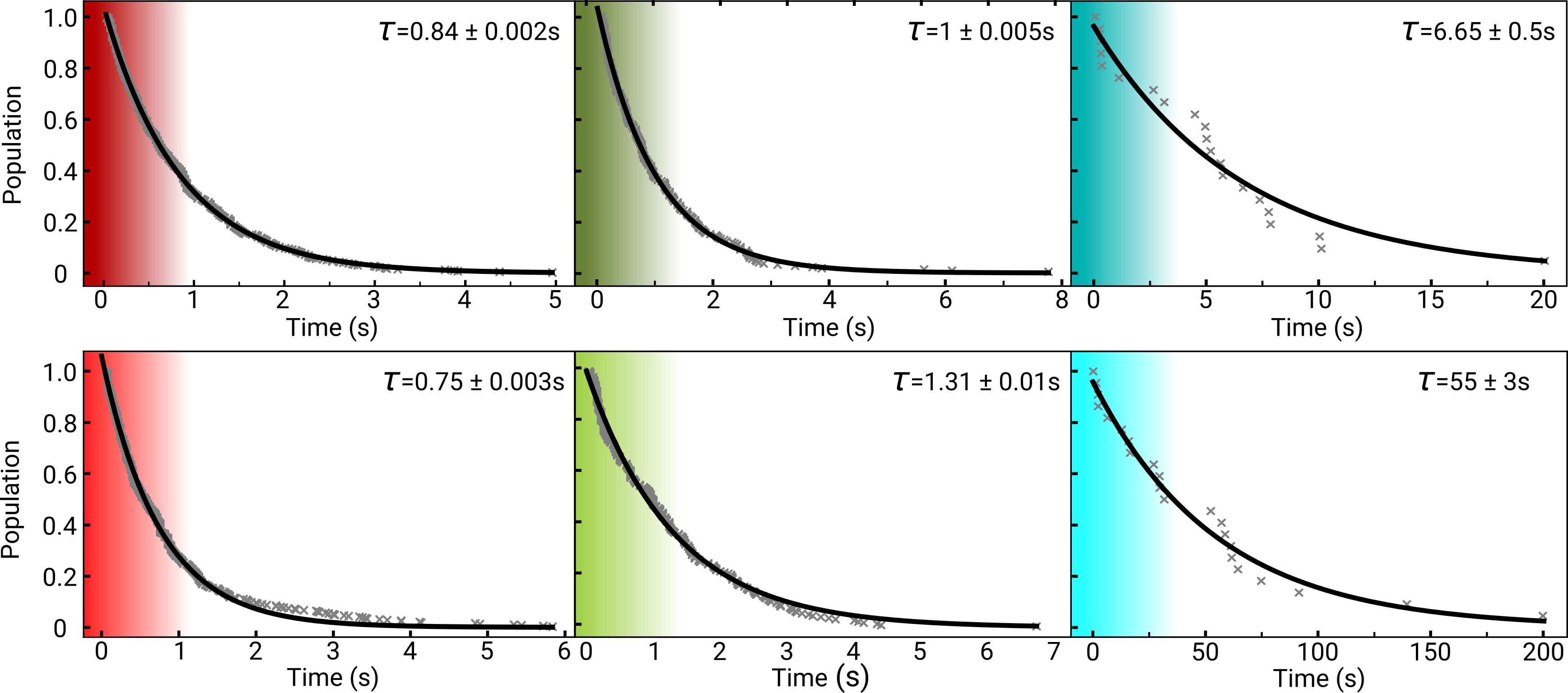}
    \caption{Exponential decay of the states representing the lifetime. For each colored section the population for upper and lower state over time was plotted. }
    \label{S_exp}
\end{figure*}
To analyze the jumps in z-position over time for the measured signal, we divided the signal in three colored sections where each has an upper and lower state. 
For each state we determined the dwell times defined as the duration for which the system remains in this state before it jumps to the corresponding level. 
A jump was counted when the signal stayed longer than two consecutive data points below the midpoint value in this section. 
Transitions were not evaluated across section boundaries. 
From these dwell times, the population fraction was evaluated assuming that the remaining population after the $i$-th event dropped to $(N-i)/N$ where $N$ represents total event count.
Subsequently, the data were fitted using an exponential decay of the form 
\begin{equation}
    f(t)= A\cdot e^{-\frac{t}{\tau}}
    \label{eq: exponential}
\end{equation}
where $\tau$ represents the mean lifetime of the respective state. 

The fitted decays with the corresponding lifetimes can be seen in figure \ref{S_exp}. 
The red region has the fastest switching rate and a nearly balanced population, though the upper state is slightly more occupied. 
This behavior suggests vicinity to an avoided level crossing, a claim supported by lifetime analysis. 
Both states decay exponentially with mean lifetimes below one second.
The population asymmetry is also reflected in the upper state's longer lifetime. 
In the green area, the switching rate is lower and the lower state shows a higher, but still similar, population.
Similar behavior is displayed by the lifetime fits, which show exponential decay and mean lifetimes slightly above \qty{1}{s}, with the more populated lower state having a slightly longer lifetime.
The slowest switching rate and a strong asymmetry toward the lower state are shown by the blue region.
Despite the reduced number of events, an exponential decay model can still be fitted.
The mean lifetimes are $\tau=\qty{6.65}{s}$ for the upper state and $\tau=\qty{55}{s}$ for the lower state, which is consistent with the observed population imbalance.
In general, an increase in population asymmetry is associated with an increase in lifetime.

\section{Effective Spin Hamiltonian parameters}
The following is a summary of the parameters used in the simulations and their sources. The anisotropy constants for the ligand field Hamiltonians of $^{163}$DyPc$_2$ were obtained from private communication with Eufemio \cite{moreno2021_personal}. The hyperfine and quadrupole parameters A and P were taken from the literature \cite{moreno2017nuclear} and converted from the EasySpin convention to the Hamiltonian definition given by equation \ref{equ1}.\\
\begin{table}[h]
\centering
\caption{Hamiltonian parameters of $^{163}$DyPc$_2$}
\begin{tabular}{c | c}
\hline
Parameter & Value (cm$^{-1}$) \\
\hline
$B_2^0$ & $376.02 \cdot (-2/9/35)$ \\
$B_2^2$ & $0.024 \cdot (-2/9/35)$ \\
$B_4^0$ & $-142.7 \cdot (-8/11/45/273)$ \\
$B_4^4$ & $2.92 \cdot (-8/11/45/273)$ \\
$B_6^0$ & $25.77 \cdot (4/11^2/13^2/3^3/7)$ \\
$B_6^4$ & $2.35 \cdot (4/11^2/13^2/3^3/7)$ \\
$A$ & $0.0051$ \\
$P$ & $0.00467$ \\
\hline
\end{tabular}
\end{table}
These parameters have been reported in the literature or the $^{159}$TbPc$_2$ system. The ligand field parameters and Stevens factors were taken from \cite{Thiele2016} and the hyperfine and quadrupole parameters, A and P, from \cite{ishikawa2005quantum}.
\begin{table}[h]
\centering
\caption{Hamiltonian parameters of $^{159}$TbPc$_2$}
\begin{tabular}{c | c}
\hline
Parameter & Value (cm$^{-1}$) \\
\hline
$B_2^0$ & $414 \cdot (-1/99)$ \\
$B_4^0$ & $-228 \cdot (2/16335)$ \\
$B_4^4$ & $100 \cdot (2/16335)$ \\
$B_6^0$ & $33 \cdot (-1/891891)$ \\
$A$ & $0.0173$ \\
$P$ & $0.01$ \\
\hline
\end{tabular}
\end{table}
\section{Random Telegraphic Noise in TbPc$_2$}
Based on the results of $^{163}$DyPc$_2$ further measurements with TbPc$_2$ were conducted. 
Terbium possesses only one naturally occurring stable isotope $^{159}$Tb$^{3+}$ which also possesses nuclear spin $I=3/2$, a prerequisite for the hyperfine analysis presented here.
By contrast, dysprosium occurs naturally as a mixture of isotopes, only some of which possess nuclear spin.
The $^{159}$Tb isotope has nuclear spin $I=3/2$, yielding four hyperfine sublevels instead of six and thereby reducing the number of overlapping level crossings. 
\begin{figure*}
    \centering
    \includegraphics[width=\textwidth]{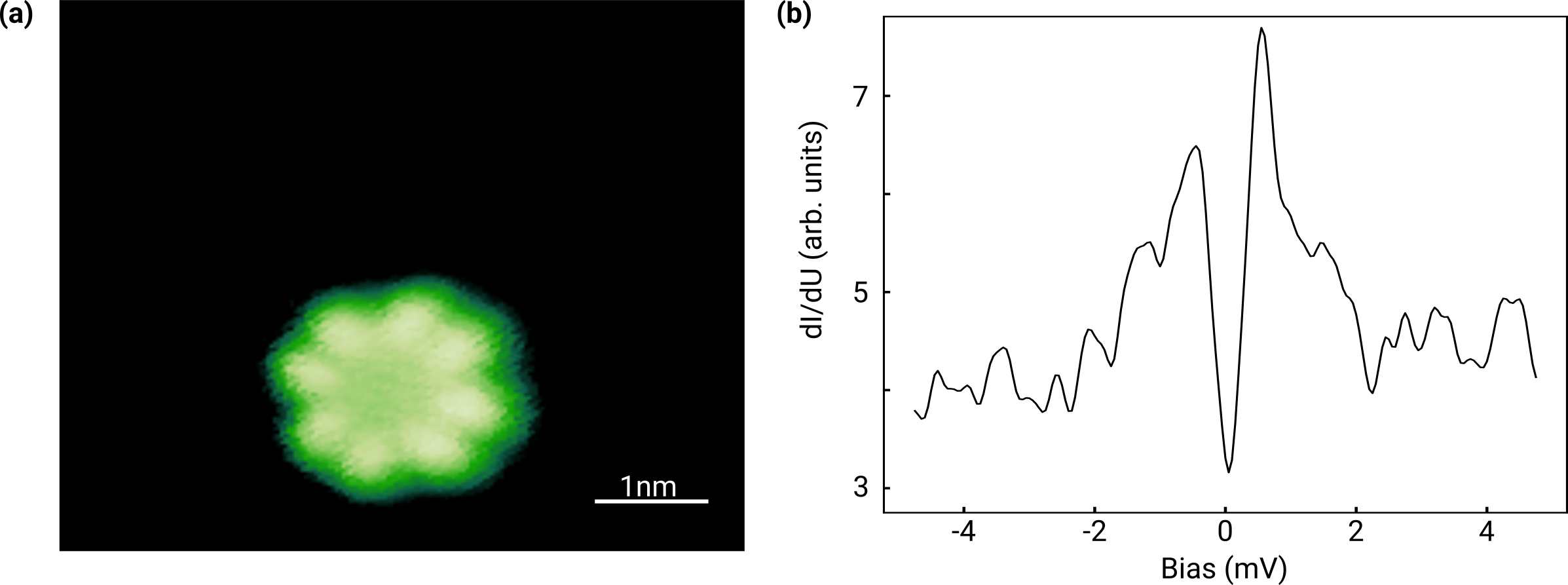}
    \caption{(a) An STM image (\SI{15}{pA}, $\SI{1}{V}$) showing TbPc$_2$ molecule adsorbed on a Au(111) surface. (b) STS spectrum (\SI{5}{mV}, \SI{8}{pA}, Lockin: $\SI{200}{\mu V}, \SI{4190}{Hz}$) acquired on one of the lobes of the lobes of the TbPc$_2$ molecule.} 
    \label{fig:TbPc2_scan}
\end{figure*}
TbPc$_2$ molecules were prepared following a procedure similar to that used for DyPc$_2$, with minor adjustments to the deposition conditions. 
On the surface, TbPc$_2$ molecules exhibit a cloverleaf structure similar to DyPc$_2$, with an apparent height of \qty{395}{pm}.
RTN measurements were performed by first characterizing the Kondo resonance symmetrically split around zero bias, exhibiting a splitting of \qty{1.02}{mV} as displayed in Fig. \ref{fig:TbPc2_scan}.
Subsequently, time traces were acquired at a bias voltage corresponding to one of the split Kondo peaks.
The corresponding results are shown in \ref{fig:TbPc2_RTN}. 
We acquired RTN data over a magnetic-field range from \qty{-100}{mT} to \qty{4}{mT} and compared the observed switching behavior. 
Figure \ref{fig:TbPc2_RTN}a presents the Zeeman diagram for TbPc$_2$ together with selected magnetic fields used for representative RTN measurements.
The yellow region (Fig.~\ref{fig:TbPc2_RTN}d) lies well separated from quantum-state crossings, and the signal remains stable.
For the green region (Figure \ref{fig:TbPc2_RTN}c) the magnetic field is close to only one crossing and sufficiently separated others, resulting in switching between two discrete states. 
In the red region (Fig.~\ref{fig:TbPc2_RTN}b), we observe three distinct sections, each exhibiting two-level behavior. 
Comparison of the complete set of time traces reveals overall agreement between experiment and theoretical predictions.

\begin{figure*}
    \centering
    \includegraphics[width=\textwidth]{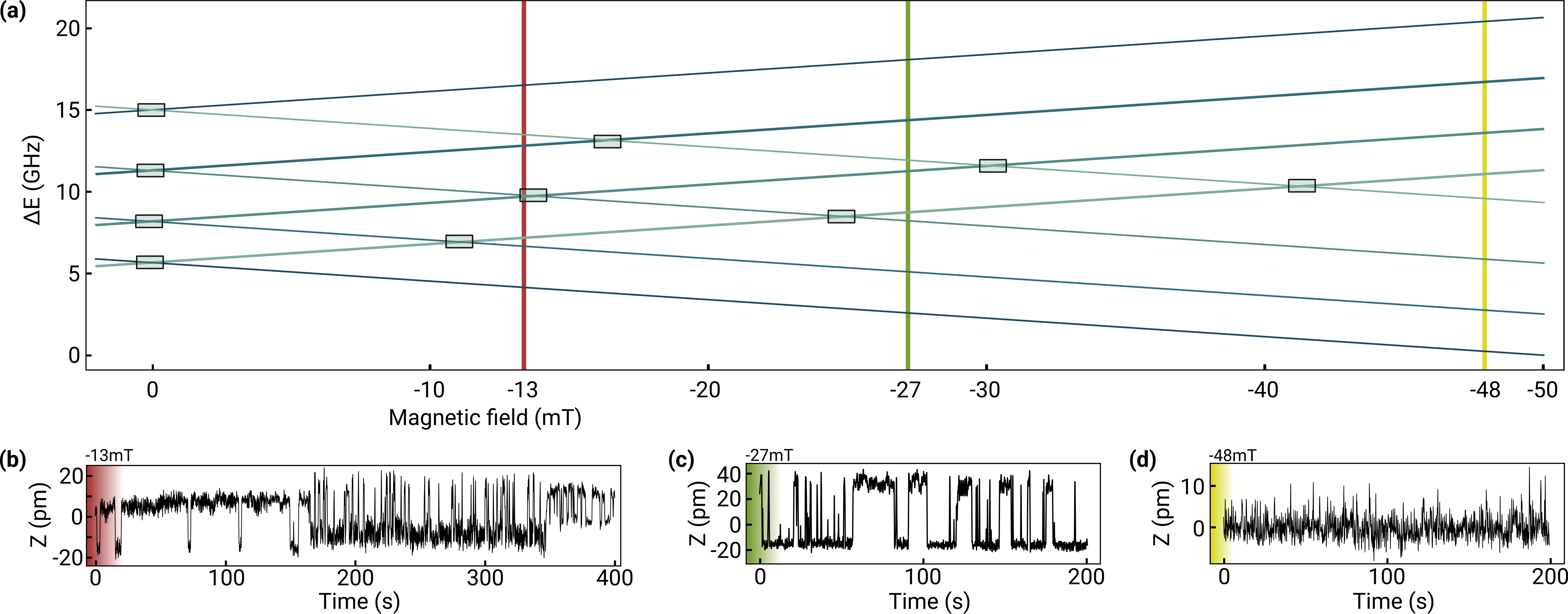}
    \caption{Magnetic field dependent RTN in TbPc$_2$ molecules. (a) Zeeman diagram for TbPc$_2$ molecule shown for reversed magnetic field orientation”. The rectangular boxes highlight all quantum-state crossings. Colored lines mark the magnetic fields corresponding to the RTN data in (b–d). (b)-(d) show representative time traces for each corresponding magnetic field with six levels distributed across three regions (b), two levels in one area (c) and no jumps between levels in (d).} 
    \label{fig:TbPc2_RTN}
\end{figure*}

\section{Third order perturbation theory}

The Kondo effect for DyPc$_2$ was modeled based on third order perturbation theory following the concepts of Ternes \cite{ternes2015spin}. Based on the Hamiltonian, transition rates between all  states induced by electron scattering from the leads were estimated for magnetic fields away from the avoided level crossings, i.e. for vanishing external field for Dy.
As the eigenstates are nearly pure in the $z$-components of the spins, spin flips are mainly induced in the ligand spin. Transitions of the nuclear spin were at least 7 orders of magnitude lower.  
Thus, the third order perturbation theory was carried out taking into account only the states with high transitions rates making the task computationally feasible.
Besides the matrix elements of the involved transitions, also the energy of the involved states enter the calculation. These include the Coulomb repulsion $U$, which does not depend on the nuclear spin, and the singly occupied state energy $\epsilon$, which does depend on the nuclear spin. 
We refrained from using a full renormalization group calculation of the Anderson impurity, as the Dy moment splits the Kondo peak substantially and the peaks can be described as inelastic excitations \cite{ternes2015spin}.

\section{General Synthetic Remarks}
All reagents and solvents (analytical grade) were purchased from commercial suppliers (Sigma-Aldrich) and used as received unless otherwise stated.

Isotopically enriched dysprosium oxide starting material, ($^{163}$Dy$_2$O$_3$ (\SI{94.6}{\%}) was purchased from BuyIsotope, Sweden. The oxide lanthanide was then converted to the acetylacetonate counterpart employing published procedures \cite{moreno2017nuclear,stitesrareearth}. 

All manipulations were conducted using standard laboratory practices.

\begin{figure*}
    \centering
    \includegraphics[width=0.7\linewidth]{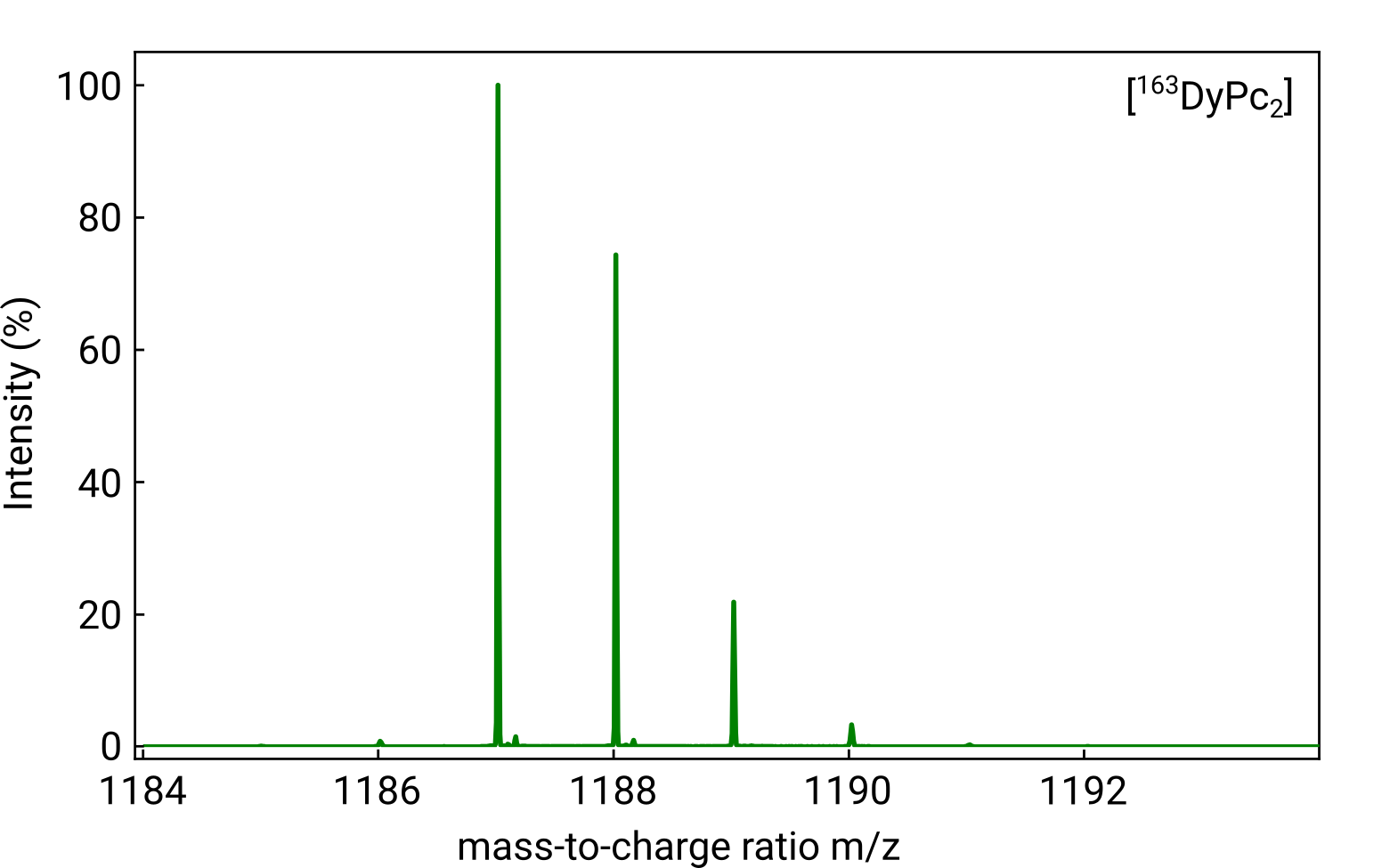}
    \caption{The positive MALDI mass spectra in a zoomed range for the intensity peak of the Isotope [$^{163}$DyPc$_2]^0$.}
    \label{sup:mass_spec}
\end{figure*}

\begin{figure*}
    \centering
    \includegraphics[width=0.7\linewidth]{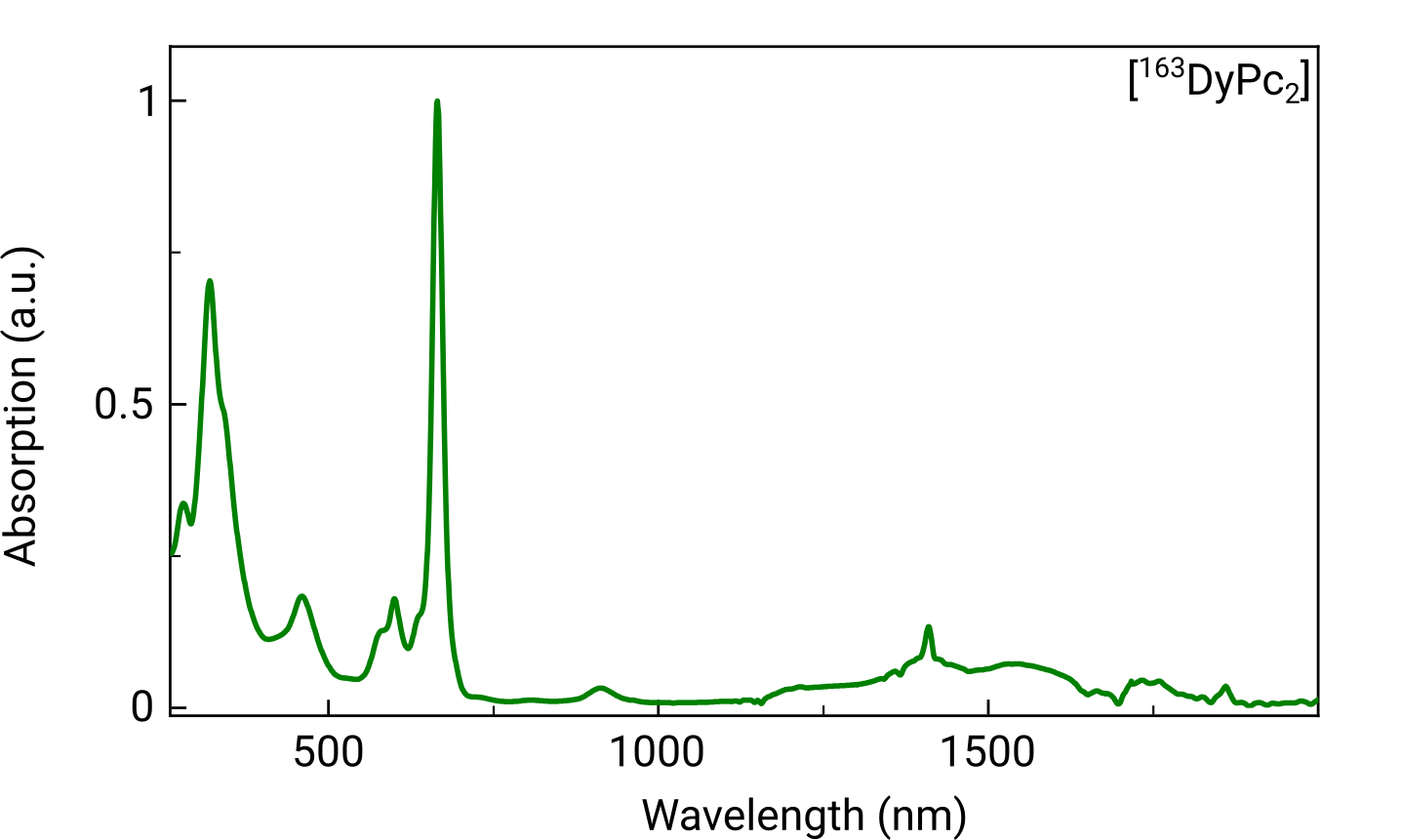}
    \caption{The UV/vis/nIR absorption spectrum of [$^{163}$DyPc$_2]^0$ is similar to the neutral green form of [LnPc$_2]^0$ spectra \cite{markovitsi1987} with the most intense band - the Q-band - at \SI{672}{nm}. In the near-infrared region two main bands are observed at about \SI{906}{nm} and \SIrange{1300}{1800}{nm}. The shorter wavelength band is related to the radical part and attributed to the 1e$_g\rightarrow$ a$_1u$ transition; the lower-energy band is assigned to an intramolecular charge transfer (CTI). Those signals are the fingerprints of the neutral species [LnPc$_2]^0$ and, thus, confirm their nature. All the near-infrared bands disappear upon reduction by hydrazine hydrate (1\% vv).}
    \label{sup:UV_spec}
\end{figure*}

\textbf{Synthesis}. The synthesis of [DyPc$_2$]$^0$ and [$^{163}$Dy$_2$]$^0$ was accomplished by templating reactions, starting from a mixture of the phthalonitrile precursor \textit{o}-dicyanobenzene and corresponding acetylacetonate Dy(acac)$_3\cdot \textit{n}$H$_2$O or $^{163}$Dy(acac)$_3\cdot \textit{n}$H$_2$O, in the presence of a strong base [e.g., 1,8-diazabicyclo[5,4,0]undec-7-ene (DBU)] and high-boiling solvents, such as pentanol according to a procedure established earlier \cite{branzolispindynamic2010,moreno2017nuclear}  and analytical data confirmed its intact structure (see Fig. \ref{sup:mass_spec} and \ref{sup:UV_spec}). By means of additional radial chromatography on silica gel followed by recrystallization from chloroform-hexane mixture, analytically pure powder samples were achieved. 

The green fraction was collected and recrystallized from chloroform/hexane to afford a black powder of [DyPc$_2$]$^0$. UV/Vis (DCM) (nm): 280, 321, 4614, 599, 665; HRMS (MALDI-ToF mass spectra for, positive mode): m/z calcd. for [M]$^+$ DyC$_{64}$H$_{32}$N$_{16}$ 1188.2400 Da; found 1188.2367 Da.
For $^{163}$Dy C$_{64}$H$_{32}$N$_{16}$: UV/Vis (DCM) (nm): 280, 321, 4614, 599, 665; 
HRMS (MALDI-ToF mass spectra for, positive mode): m/z calcd. for [M]$^{+}$ 1186.97424 Da; found 1187.02006 Da.

\textbf{Characterization} UV-visible-NIR spectra were measured on a Varian Cary 500 Scan UV/vis/NIR spectrophotometer by using a \SI{1}{cm} optical path length quartz cell. MALDI-ToF spectra were recorded on a Waters SYNAPT$^\text{TM}$-G2-Plattform in the positive ion mode.


\bibliographystyle{apsrev4-2.bst}

\bibliography{bib}
\end{document}